\documentstyle[prd,floats,aps,epsf]{revtex}

\newcommand{\beq}{\begin{equation}}
\newcommand{\eeq}{\end{equation}}
\newcommand{\beqa}{\begin{eqnarray}}
\newcommand{\eeqa}{\end{eqnarray}}

\begin{document}
\draft

\twocolumn[\hsize\textwidth\columnwidth\hsize\csname @twocolumnfalse\endcsname

\title{Binary Black Hole Mergers in 3d Numerical Relativity}
\author{Bernd Br\"ugmann 
 \\ Max-Planck-Institut f\"ur Gravitationsphysik 
 \\ Albert-Einstein-Institut
 \\ Schlaatzweg 1, 14473 Potsdam, Germany 
 \\ bruegman@aei-potsdam.mpg.de
}
\date{August 17, 1997; revised September 10, 1998}
\maketitle

\begin{abstract}
  The standard approach to the numerical evolution of black hole data
  using the ADM formulation with maximal slicing and vanishing shift
  is extended to non-symmetric black hole data containing black holes
  with linear momentum and spin by using a time-independent conformal
  rescaling based on the puncture representation of the black holes.
  We give an example for a concrete three dimensional
  numerical implementation. The main result of the simulations is that
  this approach allows for the first time to evolve through a brief
  period of the merger phase of the black hole inspiral.
\end{abstract}

\vskip2pc]
\narrowtext


\section{Introduction}

While the binary black hole problem is essentially unsolved, there
have been many advances in our understanding of the general
relativistic evolution problem.  The situation of interest is how two
black holes orbit each other, spiral inwards, eventually merge, and
emit gravitational waves in the process. Gravitational wave detectors
are being built that eventually will detect waves from such sources
\cite{detect}. Although at least as interesting from an astrophysical
as opposed to purely general relativistic view point, here we will not
consider neutron stars or other matter sources (e.g.\ \cite{matter}).
To name just three areas of research related to the inspiral of two
black holes: post-Newtonian methods are applicable if the system is
not too strongly general relativistic \cite{post-newt}, that is in the
early phase of the inspiral; perturbation methods can model the final
stages of the merger \cite{perturb}; and full numerical relativity can
be applied to the late strong-interaction phase if one imposes
axisymmetry \cite{Sm79,axisym}.

What may appear surprising is that the general, three-dimensional two
black hole problem has not been solved {\em numerically} to some
degree.  After all, there has been a lot of theoretical work on the
Einstein equations, and even though they are complicated, one might
expect that eventually computers become powerful enough to at least
allow some rough investigations into the general problem. 

However, general relativity has at least two characteristic features,
gauge freedom and singularity formation, which would prevent us from
solving the problem completely with current computer codes even if an
infinitely fast computer became available. Here gauge freedom refers
to the freedom to choose coordinates, which in numerical relativity is
a hard problem since in general one cannot specify coordinates for the
entire spacetime in advance. For an adequate representation of the
spacetime on a numerical grid one usually has to construct the
coordinates numerically as one of the steps in the numerical evolution
scheme. In this dynamic construction of coordinates one has to avoid
the formation of coordinate singularities, but even if these are absent, it
is still possible that completely regular initial data develops
physical singularities, as is the case for typical black hole spacetimes.
Both problems, the choice of coordinates and the representation of black hole
spacetimes on a numerical grid, are still awaiting a completely
satisfactory solution.

What has been achieved in 3d numerical relativity is the evolution of
weak gravitational waves \cite{wa3d1,wa3d2} (non-linear waves that are
weak enough to not collapse to black holes), and the evolution of a
single spherically symmetric black hole \cite{ss3d,BrAM,daues} in
Cartesian coordinates and for non-trivial slicings.  Recently, long
term stable evolutions have been achieved for single black holes
\cite{stable}.  A single moving black hole has been simulated in
\cite{move}, which should become important for black hole collisions.
Axisymmetric collisions of two black holes with a 3d code have been
reported in \cite{mi3d}.  Non-axisymmetric spacetimes containing a
single distorted black hole are studied in \cite{ss3ddist}.  Various
other 3d projects are actively pursued, most notably by the US binary
black hole alliance \cite{grand}.

In this paper we introduce a new approach that makes the study of
non-axisymmetric binary black hole spacetimes possible, at least for
short time intervals. The crucial new insight is that the general
binary black hole initial data of \cite{BrBr} can be evolved on $R^3$
without special inner boundary using a conformal rescaling that is
constant in time. The main result is the evolution of the apparent
horizon, which for appropriate black hole initial data has two
components initially that are replaced by a single component during
evolution.

The main limitation of the current implementation is that only a brief
interval of the merger phase of the inspiral of two black holes can be
studied. This is due to the well-known problem of grid-stretching
associated with our particular gauge choice, maximal slicing and
vanishing shift. We outline below how the gauge conditions can be
generalized. Here we want to demonstrate our approach with one minimal
example for general black hole data, a more detailed study is
left to a future publication.

To summarize, initial data for two black holes of unequal mass that
are moving and spinning are constructed following \cite{BrBr}. The
evolution is carried out in the 3+1 ADM formulation with maximal
slicing and vanishing shift using BAM, a bifunctional adaptive mesh
code for coupled elliptic and hyperbolic systems (only a fixed mesh
refinement of nested grids is used here; see \cite{BrAM,BrBr,BrBAM}
about earlier versions of this code). An evolution time of about
7--10$M$ (in units of the ADM mass) is achieved.  We can study the
geometric information in the collapse of the lapse function, and we
find the apparent horizon and show data for the transition from two
outermost marginally trapped closed surfaces to a single one.

In the following, we describe the method and its implementation, 
present the physical data, and conclude with a discussion.

\section{Description of the puncture method}

We begin with the (3+1)-dimensional Arnowitt-Deser-Misner (ADM)
decomposition of the 4-dimensional Einstein equations \cite{Yo79}.
The dynamical fields are the 3-metric $g_{ab}$ and its extrinsic
curvature $K_{ab}$ on a 3-manifold $\Sigma$, both depending on space
(points in $\Sigma$) and a time parameter, $t$. The foliation of the
4-dimensional spacetime into hypersurfaces $\Sigma$ is characterized
in the usual way by a lapse function $\alpha$ and a shift vector
$\beta^a$. The Einstein equations become
\begin{eqnarray}
        \partial_t g_{ab} &=& -2\alpha K_{ab} + D_a\beta_b + D_b\beta_a, 
\label{dgdt}
\\
        \partial_t K_{ab} &=& -D_aD_b\alpha + \alpha (R_{ab} 
                              - 2 K_{ac}{K^c}_b + K_{ab} K) 
\nonumber 
\\
        && + \beta^cD_cK_{ab} + K_{ac} D_b\beta^c + K_{cb}D_a\beta^c,
\label{dKdt}
\\
        0 &=& D^b (K_{ab} - g_{ab} K),
\label{diffeo}
\\
        0 &=& R - K_{ab} K^{ab} + K^2,
\label{hamil}
\end{eqnarray} 
where $R_{ab}$ is the 3-Ricci tensor, $R$ the Ricci scalar, $K$
the trace of the extrinsic curvature, and $D_a$ the covariant
derivative derived from the 3-metric. 

We solve the constraints on an initial hypersurface at $t = 0$ for
conformally flat two black hole data by the method introduced in
\cite{BrBr}.  The solution is found in terms of conformally
transformed quantities, $g_{ab} = \psi^4 \delta_{ab}$ and $K_{ab} =
\psi^{-2} \bar K_{ab}$.  On a Cartesian grid, one chooses two points
$\vec{c_1}$ and $\vec{c_2}$ that represent the internal asymptotically
flat regions of the two black holes.

The diffeomorphism constraint is solved by the Bowen-York extrinsic
curvature \cite{Yo89} with parameters $\vec{P}_1$ and $\vec{P}_2$ for
the linear momenta and $\vec{S}_1$ and $\vec{S}_2$ for the spins,
\begin{eqnarray}
   \bar K^{ab}_{PS} &=& \sum_{i=1}^2 (
   [\frac{3}{2r^2} (P^a n^b + P^b n^a - (\delta^{ab} - n^a n^b) P^c n_c)]_i 
\nonumber\\
   && + [\frac{3}{r^3} (\epsilon^{acd} S_c n_d n^b 
                      +\epsilon^{bcd} S_c n_d n^a)]_i),
\label{KPJ}
\end{eqnarray}
where $n^a$ is the radial normal vector in Cartesian coordinates, $n^a
= x^a/r$, and where $[\ldots]_i$ denotes the term for the $i$-th black
hole with $r_i = |\vec{x}-\vec{c_i}|$.
The vanishing of the Hamiltonian constraint $H$, (\ref{hamil}), 
becomes a non-linear elliptic equation for 
the conformal factor $\psi$. The novel feature of \cite{BrBr} is that the
equation for $\psi$ is rewritten for $R^3$, where the points 
$\vec{c_1}$ and $\vec{c_2}$ are part of the computational domain.
The solution has the form
$\psi = u + \sum_{i=1}^2 m_i/(2r_i)$,
where $u$ is an everywhere regular function. Recall that
Schwarzschild initial data can be written as $\bar K_{ab} = 0$ and
$\psi = 1 + m/(2r)$. 

We presented the initial data in some detail because this particular
construction leads to a natural treatment of the inner boundary during
evolution. On the initial slice the metric diverges at $\vec{c_1}$ and
$\vec{c_2}$ (in the right manner to represent the inner asymptotically
flat regions).  As already done in \cite{ss3d,BrAM}, an evolution of
Schwarzschild initial data can be performed on $R^3$, if the metric is
divided by the time-independent conformal factor $\psi^4 = (1 +
m/(2r))^4$, $g_{ab} = \psi^4\tilde g_{ab}$. The derivatives of the
physical metric $g_{ab}$ are obtained as the numerical derivatives of
the regular conformal metric $\tilde g_{ab}$ plus terms containing the
derivatives of the conformal factor $\psi$, which are computed
analytically. Furthermore, the grid is offset so that the puncture is
not one of the grid points.

For binary black hole data constructed as above, which (i) has a
conformal factor analogous to Schwarzschild at each puncture, and
which (ii) has a $1/r$ fall-off in the extrinsic curvature as in
(\ref{KPJ}), we find that the evolution can also be performed on
$R^3$, if the metric is rescaled with the time-independent
conformal factor $\phi = 1 + \sum_{i=1}^2 m_i/(2r_i)$.  We use $1$
instead of $u$ so that the derivatives of $\phi$, which we need to
compute the derivatives of the unscaled metric $g_{ab}$, can be
computed analytically. 

In the numerical implementation, we do not rescale the extrinsic
curvature but again place the grid such that the $\vec{c_i}$ fall
between the points of the grid.  Note that this is unproblematic for
the evolution equation (\ref{dKdt}) only if the shift vector vanishes,
because then no derivatives of the extrinsic curvature are needed, so
let us turn to the choice of lapse and shift.

To complete the specification of the evolution problem, we have to
choose a slicing condition \cite{SmYo}.  Not only can lapse and shift
be specified freely, but quite non-trivial choices have to be made,
since simple ``Cartesian'' choices like geodesic slicing ($\alpha =
1$, $\beta^a = 0$) do not work in general.  Here we choose vanishing
shift, $\beta^a = 0$, and maximal slicing for the lapse, where
$\alpha$ is the solution of 
\beq 
\Delta \alpha = \alpha K_{ab} K^{ab}.
\label{maxslice}
\eeq
Below we discuss a few generalizations. 

The question arises whether (\ref{maxslice}) is well-defined near the
punctures. The principal part of $\Delta \alpha$ is of order $r^4$,
while the terms involving first derivatives of $\alpha$ have
coefficients of order $r^3$, and $K_{ab} K^{ab} = \psi^{-12}
\delta^{ac} \delta^{bd} \bar K_{ab} \bar K_{cd}$ for puncture data is
of order $r^6$ ($r^8$ if there is no spin).  Heuristically, this
implies that the first derivatives of $\alpha$ have to vanish at the
punctures.  In the numerics, we multiply equation (\ref{maxslice}) by
$\phi^4\sim 1/r^4$ so that the principal part is of order 1, and it
turns out that for grids staggered about the punctures the first
derivatives of $\alpha$ are sufficiently close to zero near the
punctures.

Factoring out the singular behaviour in the evolution equations and in
the maximal slicing equation by factoring out the singular part of the
conformal factor of the initial data would not be useful if during the
course of the evolution the character of the singularity at the
punctures could change or if additional singularities could arise.
Suppose that $\alpha \rightarrow 1$ sufficiently fast and that
$\beta^a = 0$. From (\ref{dgdt}) and (\ref{dKdt}), near the punctures
$\partial_t g_{ab} = O(1/r)$, and also $\partial_t K_{ab} = O(1/r)$,
since $ - 2 K_{ac}{K^c}_b + K_{ab} K = O(r^2)$, but for 
$g_{ab} = (1+m/(2r))^4 \delta_{ab}$ one finds that 
$R_{rr} = - 8m/(r(m+2r)^2)$.
However, with the time-independent rescaling $g_{ab} = \psi^4\tilde g_{ab}$ 
and $K_{ab} = \psi^4\tilde K_{ab} = \psi^{-2} \bar K_{ab}$, one obtains
$\partial_t \tilde g_{ab} = 0$ and $\partial_t \tilde K_{ab}
= 0$ at the punctures. In the numerics, we still rely on the
staggering of the grid instead of a rescaling of $K_{ab}$, although
the latter may have some advantages.

The bottom line for the puncture method for evolutions using maximal
slicing with vanishing shift is that we can treat not only the initial
data but also the maximal slicing and evolution equations on a domain
without special inner boundary. As we discuss below, the numerical
implementation shows some remnants due to finite difference problems
near the punctures, but the general setup appears to be valid. Let us
add that at the outer boundary a Robin boundary
condition is used for the initial data, for the evolution and maximal
slicing the data at the outer boundary is kept fixed (cmp.\ \cite{ss3d}).

\section{Implementation and code validation}

The computer code, BAM, is a combination of a leapfrog evolution code
\cite{BrAM} and a multigrid elliptic solver \cite{BrBAM}.  One of the
drawbacks of elliptic slicing conditions in 3d codes is that they can
account for most of the computations \cite{ss3d}.  For our runs, the
elliptic solves account for 60\% of the runtime. BAM is outer loop
parallelized using Power C on an Origin 2000 (distributed shared
memory model). A typical run described below can be performed on 16
processors in 1--5 GByte of memory in 2--20 hours.

One of the interesting technical features of BAM is that it is the
first code used in 3d numerial relativity to support fixed mesh
refinements. In 3d calculations it is an enormous, and perhaps crucial
resource savings to, for example, use 4 nested cubical boxes centered
at the origin of width 8, 16, 32, and 64, each with the same number of
$V$ grid points ($4V$ points total) but with correspondingly
increasing grid spacing, instead of using a single big box of size 64
with $128\times 4V$ grid points to obtain the same resolution at the
center. Such fixed mesh refinements make it possible to move the outer
boundary sufficiently far into the $1/r$ fall-off region.  The code is
capable of (dynamical) adaptive mesh refinement \cite{BrAM}, but
currently the numerical problems at the grid interfaces and problems
with parallelization outweigh the potential benefits.

Both the ADM part and the elliptic part of BAM have been tested
separately for convergence and the propagation of the constraints, and
compared with analytic or linearized results \cite{BrAM,BrBr}.
Corresponding tests have been performed for the combined code, in
particular it reproduces the results on maximal slicing in 3d of
\cite{ss3d}. What turned out to be an important consistency and
convergence check, for a given set of parameters the code is always
run for different resolutions, different refinements and different
outer boundaries.  Since code validation is crucial, we will discuss
in the remainder of this section various issues related to
convergence.

\begin{figure}
\epsfxsize=8.5cm
\hspace{-0.0cm}
\epsffile{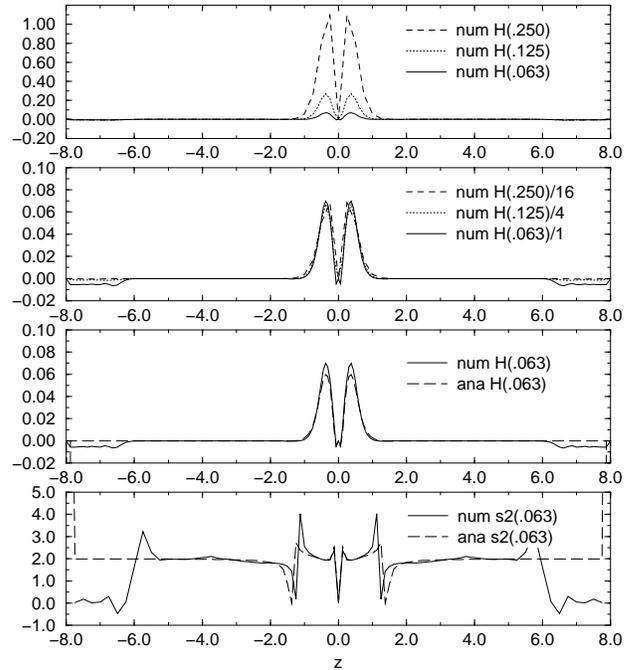}
\vspace{-0.0cm}
\caption{Convergence results for geodesically sliced Schwarzschild at 
  $t = 2$. The discretized Hamiltonian $H(h)$ is shown for both numerical 
  and analytical data.}
\label{conv1}
\end{figure}

Fig.\ \ref{conv1} shows convergence results for the ADM evolution in
geodesic slicing for a single Schwarzschild black hole of mass one.
After a finite proper time $t = \pi$, the initial slice reaches the 
physical singularity. Here we show various plots for the Hamiltonian
constraint at $t = 2$.
There are two boxes covering the cubical intervals $[-4,4]^3$ and
$[-8,8]^3$ with a refinement factor of 2. Three different central
resolutions are considered, $0.250$, $0.125$, and $0.0625$,
corresponding to $33^3$, $65^3$, and $129^3$ points per box,
respectively. 

Second order convergence in the Hamiltonian constraint $H$ is obtained
when halving the grid spacing reduces the Hamiltonian constraint by a
factor of four, $H(h) = H(2h)/4$. Assuming that a function $f$ can be
approximated at any grid point and for any grid spacing $h$ by $f(h) =
f + h^s e(h)$, where $e(h)$ is a smooth error term that is of order one in
$h$, we can use the standard two- and three-level formulas to estimate
the non-local and local order of convergence,
\beqa
  s_2(h) &=& \log_2 |f(2h)|/|f(h)|, \\
  s_3(h) &=& \log_2 |f(4h)-f(2h)|/|f(2h)-f(h)|, \\
  s_2(h,x) &=& \log_2 f(2h,x)/f(h,x), \\
  s_3(h,x) &=& \log_2 (f(4h,x)-f(2h,x)) \nonumber \\
           && \quad\quad /(f(2h,x)-f(h,x)),
\eeqa
where the two-level formulas assume that $f$ is analytically zero. We
use the $l2$-norm 
$|f(h)| = (\frac{1}{N}\sum_{i=1}^N f(h,x_i)^2)^\frac{1}{2}$.

In Fig. \ref{conv1}, we also compare the numerical results with the
analytically known solution \cite{BrAM,typo}. Of course, the finite
difference formula for $H(h)$ computed on analytic data is not
identically zero, but rather reflects the non-vanishing truncation
error of the discretization. Note that the effect of the fixed outer
boundary is clearly visible, reducing convergence to zero in a region that
actually moves inwards at about the speed of light. The inner region is
causally disconnected from this boundary problem and shows rather perfect
second order convergence, except right at the center, and at points
where the logarithm in $s_2$ becomes ill-defined. Note the small bump due
to the box nesting interface.

We now turn to the results obtained with BAM for the two black hole
problem. Obviously, there is a huge parameter space to explore, here
we restrict ourselves to one particular data set, with the overall
scale set by $m_2$: $m_1=1.5$, $m_2=1$, $\vec{c}_{1,2}=(0,0,\pm 1.5)$,
$\vec{P}_{1,2}=(\pm 2,0,0)$, $\vec{S}_1=(-0.5,0.5,0)$,
$\vec{S}_2=(0,1,1)$. The initial ADM mass estimate is $3.1$.

\begin{figure}
\epsfxsize=8.0cm
\vspace{-0.0cm}
\epsffile{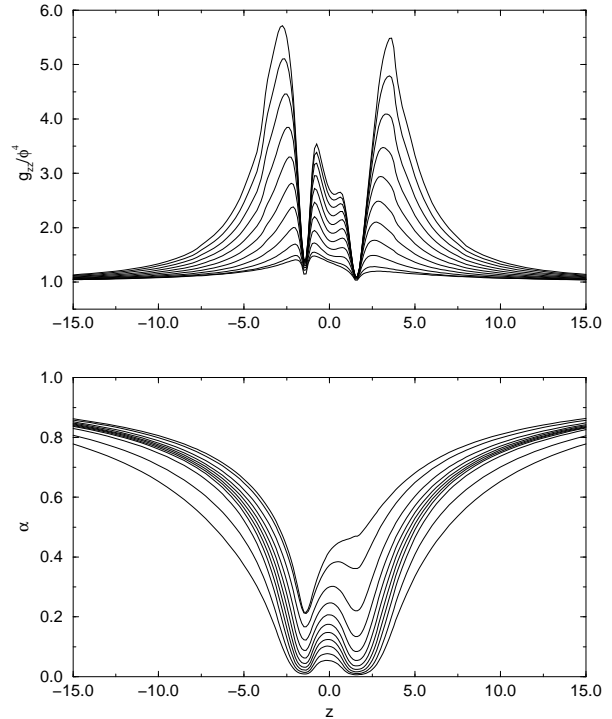}
\vspace{-0.5cm}
\caption{
Basic features of the evolution with maximal slicing and vanishing shift
for a general data set:
explosion of the rescaled metric and collapse of the lapse
  for $t = 0, 2, \ldots, 20$.}
\label{alpha}
\end{figure}

\begin{figure}
\epsfxsize=8.5cm
\hspace{0.0cm}
\epsffile{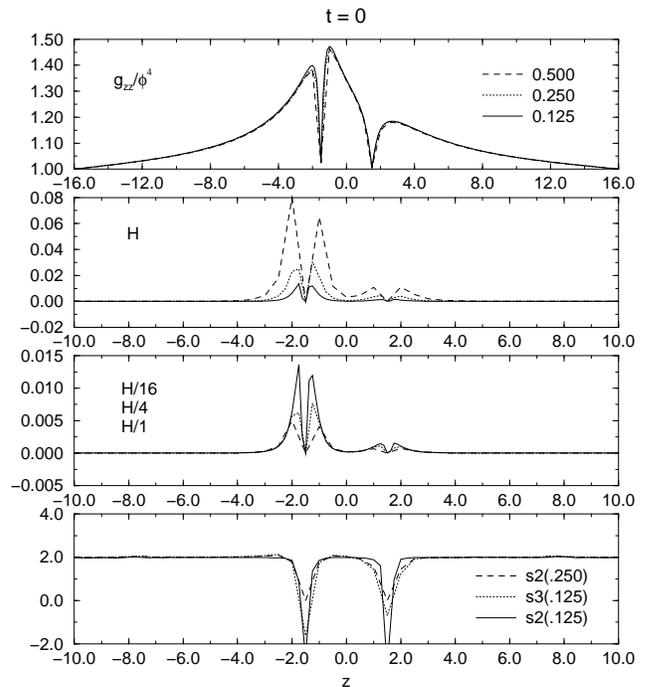}
\vspace{-0.0cm}
\caption{The rescaled metric and the Hamiltonian constraint at $t =
  0$ (general data).}
\label{conv3}
\end{figure}

The basic feature of all maximal slicing runs for black hole puncture
data is the explosion of the metric and the collapse of the lapse, see
Fig.\ \ref{alpha}. Here we use centered cubical grids with $65^3$ points
each covering $[-4,4]^3$ to $[-32,32]^3$ in spatial coordinates
with a constant refinement factor of 2. The grid spacing therefore
ranges from $0.125$ at the center to $1.000$ in the outer regions.

For these parameters, the code runs to about $t \approx 22$.  For
Schwarzschild and widely separated Brill-Lindquist holes about $t
\approx 25$--$30$ is reached, which is the same range as reported for
maximally sliced Schwarzschild in \cite{ss3d}. The characteristic
feature of maximal slicing is that the lapse collapses near the holes,
$\alpha\rightarrow 0$, while the slice advances with $\alpha=1$ at
infinity. This has been found to avoid the physical singularities for
Schwarzschild and Misner data, but the ``grid-stretching'' leads to
divergent metric coefficients.  The code breaks down first when
solving the maximal slicing equation.  The same problem occurs at
about the same time if a simple SOR (simultaneous overrelaxation)
elliptic solver is used instead of the multigrid solver, although it
is not completely clear whether this is a general problem of
(\ref{maxslice}) or of the numerics (cmp.\ \cite{Sm79}).

Fig.\ \ref{alpha} shows that for our data set the lapse collapses with
roughly the expected speed for the different masses.  After $t\approx
20$, the central region containing both holes will not evolve much
further, but as we will discuss in the next section, the apparent
horizon of the two holes keeps moving outward.

\begin{figure}
\hspace{0cm}
\epsfxsize=8.5cm
\epsffile{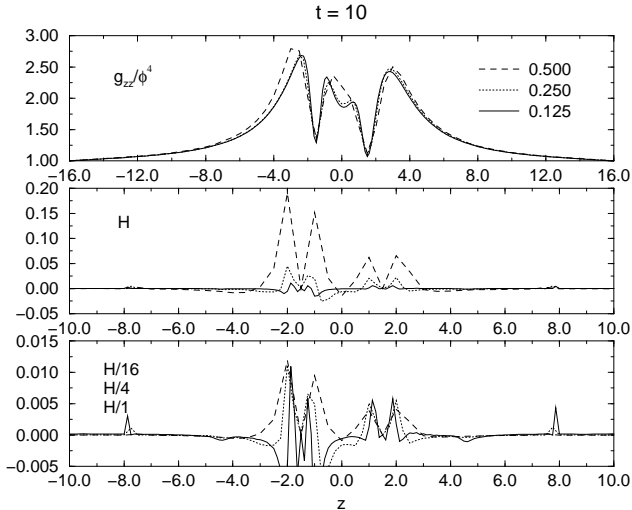}
\vspace{-2.0cm}
\caption{The rescaled metric and the Hamiltonian constraint at $t =
  10$ (general data).}
\label{conv4}
\end{figure}

\begin{figure}
\hspace{0cm}
\epsfxsize=8.5cm
\epsffile{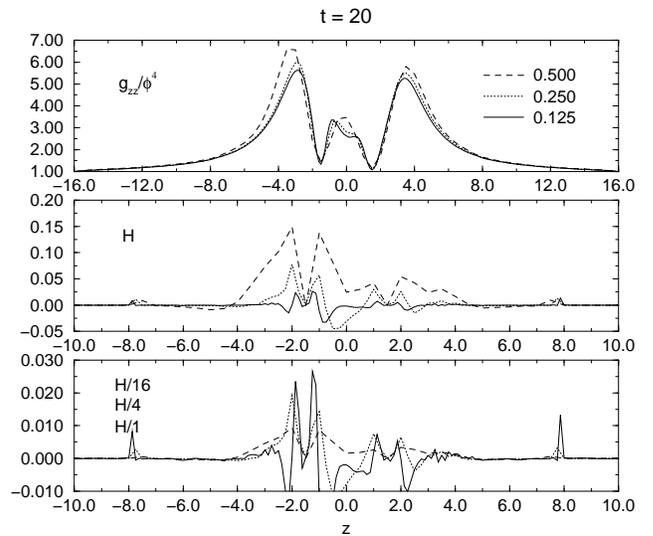}
\vspace{-2.0cm}
\caption{The rescaled metric and the Hamiltonian constraint at $t =
  20$ (general data).}
\label{conv5}
\end{figure}

\begin{figure}
\vspace{-4.0cm}
\epsfxsize=8.5cm
\epsffile{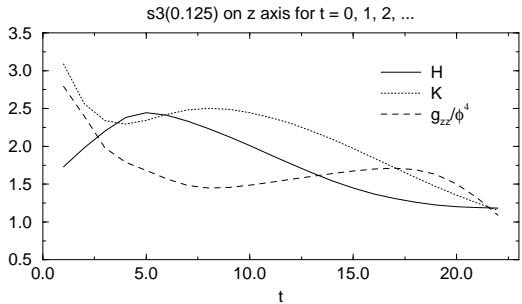}
\vspace{-0.0cm}
\caption{Three-level convergence on the $z$-axis
  versus time for the rescaled metric, the trace of the extrinsic
  curvature, and the Hamiltonian constraint (general data).}
\label{conv7}
\end{figure}

In Figs.\ \ref{conv3}--\ref{conv5} we show the rescaled metric and the
Hamiltonian constraint at times $t = 0, 10$, and $20$. In Fig.\
\ref{conv7}, the three-level convergence based on the norm over
the $z$-axis is shown during the course of the run.  This run uses two
grids covering $[-8,8]^3$ and $[-16,16]^3$.  Three different central
resolutions are considered, $0.500$, $0.250$, and $0.125$,
corresponding to $33^3$, $65^3$, and $129^3$ points per box, which is
chosen a factor of two coarser than for the Schwarzschild run in order
to move the outer boundary outwards by a factor of two.

At $t = 0$, Fig.\ \ref{conv3}, the initial data is second order
convergent except right near the punctures. As noted in \cite{BrBr},
even for rather few points across the punctures, the convergence rate
away from the punctures is not affected. When just considering initial
data, one can increase the central resolution and resolve e.g. the
rather rough humps in the Hamiltonian constraint. For evolution
problems, both the outer boundary and the inner box interfaces become
more of a problem, and therefore we had to settle on $h = 0.125$ as
the best currently achievable resolution in the center.

At later times, $t = 10$ in Fig.\ \ref{conv4} and $t = 20$ in Fig.\ 
\ref{conv5}, note that the critical region near the punctures does not
loose convergence, which remains around two for the maxima in the
Hamiltonian constraint. There is some high frequency noise visible in
the Hamiltonian constraint at $t = 20$, which is due to the inherent
mesh drifting of the double leapfrog scheme. For longer run times, it
may become important to use a more sophisticated scheme.

The spikes in $H$ at $z = \pm 8$ are due to the interpolations and
injections at the box interfaces, which are implemented as second
order or higher and should lead to at least first order in time.
Although these spikes do no seem to have a profound effect on the
evolution, they do pose a tricky problem for the following reason,
which is intrinsic to black hole evolutions with grid-stretching.  The
explosion in the metric always happens faster on the coarser, outer
grid, so that there is an inherent drift between the two grids. This
clearly needs further investigation.

Finally, consider Fig.\ \ref{conv7}. While taking the norm over the
$z$-axis smooths out all the local noise, it is still reassuring that
there is no catastrophic loss of convergence during this run. In
particular, the metric converges to at least order 1.5 until $t =
20$. In comparison, for the analytic Schwarzschild spacetime discussed
earlier, one finds convergence of only around 1.8 due to finite size effects.

Let us mention another problem that effects convergence.
Comparing with geodesic slicing, maximal slicing adds the problem
that now local errors, in particular at the outer boundary and at the
punctures, can spread instantaneously across the grid because of the
elliptic character of this gauge choice. Even though the trace of
$K_{ab}$ is roughly second order convergent, Fig.\ \ref{conv7}, we
observe drifting in the maximal slices, which have effects in the time
domain similar to the zero
crossing artifacts in the convergence rate for the Hamiltonian,
Fig.\ \ref{conv1}. 

In summary, based on the above convergence analysis, the combined
computer code is converging during evolutions at roughly an order of
1.5 or better. In particular, there is no indication that the presence
of the punctures in the domain of computation destabilizes the code,
at least for the achievable run times. The run times appear to be
limited by steep gradients occuring in maximal slicing away from the
punctures.

\section{Results and discussion}

Black hole regions of a spacetime are defined through the existence of
event horizons, which have been found numerically \cite{EHF}, but
given the available time interval it is more promising to study the
apparent horizon, which is defined for each spatial slice.  We
therefore look for a closed 2d surface $S$ whose outgoing null
expansion vanishes, $E[S] = 0$.  Solving such a generalized minimal
surface equation is an involved problem
\cite{AHF,AHF:To,AHF:Gu,AHF:cactus}, but for our purpose we found that
the following simple implementation of a curvature flow method
\cite{AHF:To} is adequate.  In terms of the standard Cartesian and
spherical coordinates, a surface which is topologically a sphere can
be parametrized by
$ u(x,y,z) = r - h(\theta, \varphi) = 0$, 
and from \cite{Yo89} with surface normal
$s^a = \partial^a u /
|\partial u|$, $|\partial u|^2 = g^{ab} \partial_a u \partial_b u$,
it follows that
\beq
E[u] = (g^{ab} - \frac{\partial^a u \partial^b u}{|\partial u|^2})
  (\frac{1}{|\partial u|} D_a \partial_b u - K_{ab}).
\label{Hexp}
\eeq
Note that we can evaluate $E$ on a 3d Cartesian grid without explicitly
introducing a 2d surface parametrized by $(\theta,\varphi)$ since 
$E^{(2)}[h](\theta,\varphi)=E^{(3)}[r-h](r = h,\theta,\varphi)
=E[u](x',y',z')$
where ${x'}^a = \frac{h}{r} x^a = (1-\frac{u}{r}) x^a$. 
Starting with a large sphere, we iterate
$u^{new} = u - d\lambda |\partial u| E[u]$ for $d\lambda$ the largest
step size that leaves the method stable.

\begin{figure}
\epsfxsize=8.5cm
\epsffile{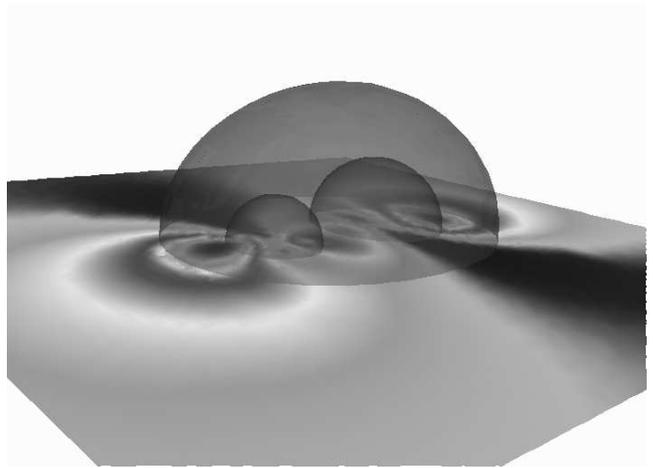}
\bigskip
\caption{Marginally trapped surfaces and $g_{zz}$ in the $y=0$ plane
  for $t = 20$, data from the two innermost nested boxes.}
\label{ah3d}
\end{figure}

\begin{figure*}
\vspace{0.5cm}
\hspace{0mm}
\epsfxsize=16cm
\epsffile{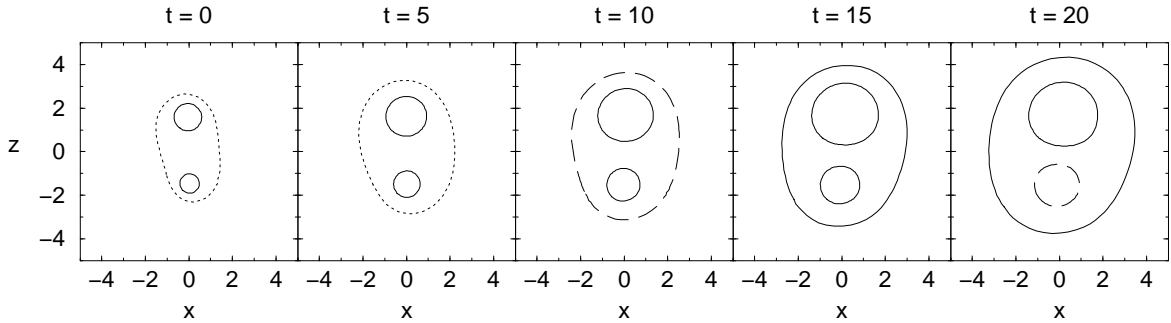}
\bigskip
\caption{Evolution of the apparent horizon. By
  construction, the dynamics is contained in the metric while the
  centers of the black holes do not move.  Solid lines indicate that a
  solution to $E = 0$ has been found, while dotted lines represent
  surfaces which are numerically close to $E = 0$ everywhere but
  analytically the flow does not stop and the existence of a true
  solution can be ruled out (based on numerical error
  estimates for Schwarzschild and Brill-Lindquist data, see also
  [25]).}
\label{ah}
\end{figure*}

Fig.\ \ref{ah3d} shows a 3d visualization of the situation at $t=20$,
while Fig.\ \ref{ah} sketches the evolution in coordinate time $t$
(same data as for Fig.\ \ref{alpha}).  Note that even if the maximal
$t$ was larger, we cannot expect to see several revolutions of two
``bodies'', and that for two reasons: the lapse freezes the evolution
in the interior, and by construction the punctures at $\vec{c}_i$ do
not move (see \cite{move} for moving single black holes).
 
Since ``freezing the evolution of the black holes'' is a limitation of
our method, we want to point out that the limitation is far less
severe than is immediately apparent.  The black holes are
characterized by a center (the puncture) and the apparent horizon. The
punctures are fixed on the grid, but they just correspond to the
internal, asymptotically flat region of the holes, not the surface of
the black holes.

The apparent horizon does move. The lapse collapses and freezes the
region around the punctures completely, but the region near the
horizon remains dynamic.  In head-on collisions of black holes
\cite{axisym} with maximal slicing it was found that the gravitational
variables are sufficiently dynamic to allow for the emission of the
gravitational waves that an observer at infinity will measure.  Given
the robustness of wave extraction in 3d demonstrated in
\cite{ss3ddist} for a single distorted black hole and a collapsing
lapse, one would expect that good wave extraction is possible for
binary black holes in 3d if run times of about $30 M_{ADM}$ can be
achieved, even if one uses maximal slicing and vanishing shift.

Allowing a non-vanishing shift can reduce the strain and the shear of
the coordinates \cite{SmYo}.  A recent proposal for a minimal strain
shift and an algebraic lapse to construct approximately corotating
coordinates can be found in \cite{Th98}.  Even if the punctures are
fixed at a coordinate location, one can code the orbital
motion in the shift vector.  Furthermore, it should be possible to
introduce a translation of the punctures through a shift vector.  Near
the punctures, the conformal factor would not change. Even when the
lapse has collapsed to zero, a non-vanishing shift leads to motion of
the punctures, see (\ref{dgdt}) and (\ref{dKdt}). 

While maximal slicing with its singularity avoing property may
be sufficient for some tasks of wave extraction, one has to look
elsewhere for long time evolutions. A simple scenario, although
technically involved, is offered by apparent horizon boundary
conditions. The puncture method with maximal slicing can be
used to start the evolution, while the apparent horizon boundary
condition takes over at a later time \cite{ss3d}.


\begin{figure}
\hspace{1.5cm}
\epsfxsize=8.5cm
\epsffile{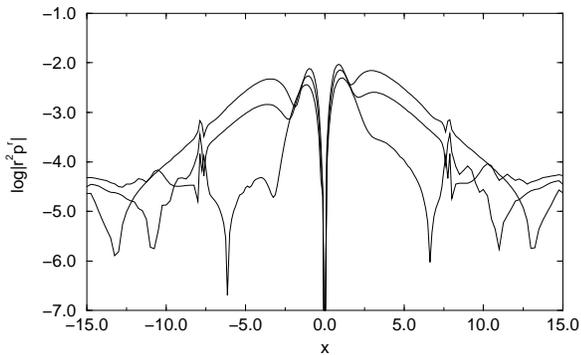}
\bigskip
\caption{Wave indicator $\log | r^2 p^r |$ along the $x$-axis at 
  $t = 9.25, 12.75, 16.25$. A ``wave'' defined by the zeros in $p^r$
  ($-\infty$ in the plot)
  propagates outwards, passing the numerical ``noise'' near the
  refinement interface at $x = \pm 8$ (cmp.\ Fig.\ 5 of [5]).
} 
\label{belrob}
\end{figure}

To find out whether the code allows wave-like
phenomena, we compute the Bel-Robinson flux, which assuming the 
Einstein equations are satisfied can be expressed as \cite{Sm79,Yo89} 
\beq 
p^c \equiv E_{ab} \epsilon^{bcd} {B_d}^a = P_{ab} Q^{cab}, 
\eeq 
where $E_{ab}$ and $B_{ab}$ are the electric and magnetic part of the
conformal tensor, and $Q_{abc}$ and $P_{ab}$ are the projections of
the 4d Riemann tensor defined in \cite{Yo89}.  A typical result is
shown in Fig.\ \ref{belrob} for two nested boxes with $129^3$ points
covering $[-8,8]^3$ and $[-16,16]^3$. Note that we do not integrate
over a sphere to obtain the total flux, which would give a much
smoother picture. Rather we plot the rescaled radial flux on the
$x$-axis to display the local order of magnitude of the signal and the
numerical noise at the interior grid faces. In a higher dimensional
plot of axisymmetric data one finds cubical shaped noise but also
rather clean axisymmetric signals.  Let us emphasize that in a binary
black hole scenario one can interpret the metric in a consistent,
gauge invariant manner as a wave travelling on a background spacetime
only in the asymptotically flat region, which is not done here. The
Bel-Robinson flux is computed only to show that wave-like phenomena
occur (in particular, they are unrelated to the location of the
apparent horizon).

\section{Conclusion}

In conclusion, we have shown how the standard approach of ADM
evolution with maximal slicing and vanishing shift can be applied to
non-symmetric black hole data containing black holes with linear
momentum and spin by using a time-independent conformal rescaling
based on the puncture representation of the black holes. We discuss an
example based on a concrete three dimensional numerical
implementation. The main result of the simulations is that this approach
allows for the first time to evolve through a brief period of the
merger phase of the black hole inspiral. Looking back to the early
numerical work of Smarr and Eppley on axisymmetric black hole
collisions in the seventies \cite{Sm79}, we feel that it is useful to
point out what concretely can be done about the two black hole problem
today, even if it is just a first step. 

Important issues for further investigations are extending the run time
so that wave extraction becomes possible, non-vanishing shift
conditions, and the implementation of an apparent horizon boundary
condition.  These issues will be addressed in a new collaborative
computational framework, the Cactus code \cite{cactus}, which provides
the infrastructure for input-output and MPI (The Message Passing
Interface) based parallelism, and which allows easy code-sharing among
several authors. For example, Cactus includes various evolution
modules, initial data set constructions, and analysis modules for wave
extraction and apparent horizon finding \cite{AHF:cactus,cactus}. In
particular, the ADM evolution routines and the multigrid elliptic
solver of BAM have already been ported to Cactus, adaptive mesh
refinement is still under development. An important and immediate
application of this new, more powerful framework will be the
comparison of the evolution of axisymmetric black hole data in 3d with
results obtained with 2d codes \cite{axisym}.


It is a pleasure to thank J. Ehlers, C. Gundlach, P. H\"ubner, B.
Schutz, E. Seidel, P. Walker, and especially S. Brandt and B. Schmidt,
for their support and many stimulating discussions. 
The computations were performed at the AEI in Potsdam.

\newcommand{\bb}{B. Br\"ugmann}
\newcommand{\bib}[1]{\bibitem{#1}}
\newcommand{\EM}{}
\newcommand{\apny}[1]{{\EM Ann.\ Phys.\ (N.Y.) }{\bf #1}}
\newcommand{\cjm}[1]{{\EM Canadian\ J.\ Math.\ }{\bf #1}}
\newcommand{\cmp}[1]{{\EM Commun.\ Math.\ Phys.\ }{\bf #1}}
\newcommand{\cqg}[1]{{\EM Class.\ Quan.\ Grav.\ }{\bf #1}}
\newcommand{\grg}[1]{{\EM Gen.\ Rel.\ Grav.\ }{\bf #1}}
\newcommand{\jgp}[1]{{\EM J. Geom.\ Phys.\ }{\bf #1}}
\newcommand{\ijmp}[1]{{\EM Int.\ J. Mod.\ Phys.\ }{\bf #1}}
\newcommand{\JCP}[1]{{\EM J. Comp.\ Phys.\ }{\bf #1}}
\newcommand{\jmp}[1]{{\EM J. Math.\ Phys.\ }{\bf #1}}
\newcommand{\mpl}[1]{{\EM Mod.\ Phys.\ Lett.\ }{\bf #1}}
\newcommand{\np}[1]{{\EM Nucl.\ Phys.\ }{\bf #1}}
\newcommand{\PL}[1]{{\EM Phys.\ Lett.\ }{\bf #1}}
\newcommand{\pr}[1]{{\EM Phys.\ Rev.\ }{\bf #1}}
\newcommand{\PRL}[1]{{\EM Phys.\ Rev.\ Lett.\ }{\bf #1}}


\end{document}